\documentclass[doublecol]{epl2} 
\usepackage[colorlinks=true,linkcolor=blue,urlcolor=blue,citecolor=blue]{hyperref}
\usepackage{amsmath}

\usepackage{mathtools}
\usepackage{mathrsfs}
\usepackage{csquotes}
\usepackage{braket}
\usepackage{amssymb}
\newcommand{\beq}{\begin{equation}}
\newcommand{\eeq}{\end{equation}}
\newcommand{\bea}{\begin{eqnarray}}
\newcommand{\eea}{\end{eqnarray}}

\title{Kinetic energy from the cubic sum rule of the dynamic structure factor}
\shorttitle{Kinetic energy from the cubic sum rule of the dynamic structure factor} 
\author{Fotios Kalkavouras\inst{1} \and Panagiotis Tolias\inst{1} \and Sebastian Schwalbe\inst{2} \and Thomas Gawne\inst{3} \and Alexander Benedix Robles\inst{2,4} \and Jan Vorberger\inst{2} \and Zhandos A. Moldabekov\inst{2} \and Maximilian P.~B\"ohme\inst{5} \and Tobias Dornheim\inst{2,3,\#}}
\shortauthor{F. Kalkavouras \etal}
\institute{                    
  \inst{1} Electromagnetics and Plasma Physics, Royal Institute of Technology (KTH), Stockholm SE-100 44, Sweden\\
  \inst{2} Institute of Radiation Physics, Helmholtz-Zentrum Dresden-Rossendorf (HZDR), D-01328 Dresden, Germany\\\inst{3} Center for Advanced Systems Understanding (CASUS), D-02826 G\"orlitz, Germany\\
  \inst{4} Technische Universit\"at Dresden, D-01062 Dresden, Germany\\
  \inst{5} Lawrence Livermore National Laboratory (LLNL), California 94550 Livermore, USA\\
  \inst{\#}Electronic mail: t.dornheim@hzdr.de
}
\abstract{The third frequency moment sum rule of the dynamic structure factor $S(\mathbf{q},\omega)$ is explored for the first time as an alternative estimator of the kinetic energy $K$ of quantum many-body systems. As a practical example, the uniform electron gas at warm dense matter conditions is considered. First, $K$ is extracted from quasi-exact \emph{ab initio} path integral Monte Carlo results for the imaginary-time density--density correlation function $F(\mathbf{q},\tau)$ and the expected excellent self-consistency with the thermodynamic differentiation route is confirmed. Second, $K$ is extracted from approximate dielectric formalism results for $S(\mathbf{q},\omega)$ and it is observed that common semi-classical approximations lead to a wave-number dependent $K$ with an incorrect short-wavelength limit. Our results are expected to be of broad interest for a great variety of applications, including time-dependent density functional theory, dielectric formalism schemes and warm dense matter models, as well as for the design of dedicated x-ray Thomson scattering experiments with the potential to provide model-free access to the full electronic equation of state.}
\begin{document}
\maketitle

\section{Introduction} The accurate description of interacting quantum many-body systems constitutes the backbone of a plethora of disciplines, encompassing as diverse a set of fields as quantum chemistry, material science, ultracold atoms, and exotic warm dense matter~\cite{quantum_theory,mahan1990many,nolting,vorberger2025roadmapwarmdensematter}.
The absence of exact analytical solutions for all but the most simple toy problems generally necessitates the application of numerical methods, which are available on different levels of sophistication and complexity. For instance, simple mean field ans\"atze are capable of capturing the correct plasmon limit of charged quantum gases~\cite{pines,pines_nozieres_bookI,bonitz_book}, but they fail to capture \emph{roton type} excitations that emerge due to electronic exchange--correlation (XC) effects~\cite{dornheim_dynamic,Dornheim_Nature_2022,koskelo2023shortrange}. The workhorse of modern electronic structure theory is given by density functional theory (DFT)~\cite{Jones_RMP_2015}, and its time-dependent extensions~\cite{marques2012fundamentals,Runge_Gross_prl_1984,dynamic1} also give access to various spectral properties. The highest rung is arguably occupied by the gamut of quantum Monte Carlo (QMC) techniques~\cite{Foulkes_RMP_2001}, which can be grouped into ground state methods (e.g., diffusion and variational Monte Carlo~\cite{anderson2007quantum}) and finite-temperature methods (e.g., path integral Monte Carlo (PIMC)~\cite{cep}).

Any meaningful comparison between methods, and in particular between theory and experiment, requires a suitable physical quantity that (a) can be readily computed, (b) can be readily observed, and (c) contains a sufficient amount of ``non-trivial'' information. In this sense, widely considered thermodynamic properties such as the pressure $p$ only fulfill (a) and (b), but their information content is rather limited, since they are essentially integrated quantities. Spectrally resolved properties constitute a potentially superior alternative because they pose more stringent requirements for theory.

In the present work, we focus on the dynamic structure factor (DSF) $S(\mathbf{q},\omega)$, which is defined as the Fourier transform of the intermediate scattering function $F(\mathbf{q},t)=\braket{\hat{n}(\mathbf{q},t)\hat{n}(-\mathbf{q},0)}$~\cite{sheffield2010plasma,hansen2013theory}. In terms of physics, we focus on the warm dense matter (WDM) regime that is particularly relevant for our understanding of astrophysical systems and for inertial confinement fusion applications~\cite{vorberger2025roadmapwarmdensematter,Ott2018}. At the same time, it should be pointed out that the DSF plays a crucial role for many other applications, including the study of ultracold atoms~\cite{Filinov_PRA_2012,Dornheim_SciRep_2022,Ferre_PRB_2016}, 2D systems~\cite{Godfrin2012}, and material science at ambient conditions~\cite{Weissker_Lifetime_TDDFT,Schuelke_Si,Sturm_IXSS_1992}. For WDM, the frequency dependence of $S(\mathbf{q},\omega)$ for a fixed momentum transfer $\mathbf{q}$ can be probed with x-ray Thomson scattering (XRTS)~\cite{siegfried_review}, which has been well-established as a standard diagnostic for extreme states of matter~\cite{Tilo_Nature_2023,Dornheim_NatComm_2025,dornheim2026overviewxraythomsonscattering,Fletcher2015}.
In particular, matching theoretical predictions of $S(\mathbf{q},\omega)$ to the experimental observation allows one to infer the nominal system parameters (e.g., mass density $\rho$, temperature $T$, $\dots$). Popular models include the comparably simple Chihara ansatz~\cite{Chihara_1987,bellenbaum2026xraydiagnosticsanalysisverification} based on the free electron gas~\cite{dornheim_dynamic} and, since relatively recently, computationally more involved \emph{ab initio} time-dependent DFT simulations~\cite{dynamic2,moldabekov2025applying,Bespalov2026}.
In practice, it has also proven useful to consider the two-sided Laplace transform of the DSF~\cite{Dornheim_T_2022,Dornheim_NatComm_2025,gawne2026modelfreeinterpretationxraythomson},
\begin{eqnarray}\label{eq:Laplace}
F(\mathbf{q},\tau) = \int_{-\infty}^\infty \textnormal{d}\omega\ S(\mathbf{q},\omega)\ e^{-\hbar\tau\omega} \, ,
\end{eqnarray}
which defines the imaginary-time analogue of $F(\mathbf{q},t)$ with $t=-i\hbar\tau$, $\tau\in[0,\beta]$ and $\beta=1/k_\textnormal{B}T$. First, transforming the XRTS signal to the imaginary-time domain allows for a stable deconvolution, which facilitates the model-free extraction of various parameters such as the temperature of the system~\cite{Dornheim_T_2022,Dornheim_T2_2022}; see Ref.~\cite{gawne2026modelfreeinterpretationxraythomson} for a topical review. Second, Eq.~(\ref{eq:Laplace}) is of key importance to the QMC community, since it connects the imaginary-time correlation function (ITCF) $F(\mathbf{q},\tau)$---readily available, e.g., from PIMC simulations~\cite{dornheim_dynamic,Dornheim_insight_2022}---to XRTS measurements.
It is noted that inversion of Eq.~(\ref{eq:Laplace}) to reconstruct $S(\mathbf{q},\omega)$ is the notoriously difficult, ill-posed \emph{analytic continuation} problem~\cite{JARRELL1996133,Chuna_JPA_2025}.

The key aim of the present work is to highlight a particularly interesting piece of information that is encoded in $S(\mathbf{q},\omega)$ and in $F(\mathbf{q},\tau)$. Being equivalent two-body correlation functions, it is intuitive that the DSF and ITCF give one access to the interaction energy $W$ provided that their entire dependence on $\mathbf{q}$ was known~\cite{Dornheim_review}. It is less intuitive that knowledge of the DSF/ITCF for a state point is sufficient to determine the kinetic energy $K$ for the same state point. This clearly implies that the full equation-of-state ($W$, $K$, $p$) can be obtained directly from the DSF/ITCF without the need for any thermodynamic integration; an observation of potential interest to theory~\cite{stls2} and experiment~\cite{Fletcher2015} alike. In addition, computing $K$ from $S(\mathbf{q},\omega)$ for different $\mathbf{q}$ values serves as an easy self-consistency check for dynamic quantum many-body theories (such as time-dependent DFT or dielectric theories) as the thermodynamic expectation value for the kinetic energy is uniquely defined and must give the same value for all wave vectors. 

In what follows, we first outline the idea in terms of the DSF and the ITCF. As a practical example, we investigate the warm dense uniform electron gas (UEG)~\cite{review}, which is often considered as the archetypal system of interacting electrons. Then, we apply the idea to quasi-exact (i.e., exact within the given Monte Carlo error bars) PIMC results for the ITCF, which indeed show the expected consistency with a level of significance of $\sim0.1\%$. After, we consider dielectric formalism schemes and obtain $K$ from the DSF; we find that common approximations such as the random phase approximation (RPA)~\cite{Gupta_1980,Perrot_1984} and the Singwi-Tosi-Land-Sj\"olander (STLS) scheme~\cite{stls_original,stls} yield a wave-number dependent kinetic energy with an incorrect short-wavelength limit, which suggests that our idea can be utilized as an extra consistency check for dielectric schemes beyond the compressibility sum rule. Finally, we discuss the relevance of our results for various applications, including DFT simulations and XRTS experiments.

\section{General theory} We shall use Hartree atomic units in the following. Let us consider the frequency moments of the dynamic structure factor, which are defined as~\cite{Dornheim_moments_2023}:
\begin{equation}
    M_S^{(\alpha)}(\mathbf{q})=\int_{-\infty}^{+\infty}\textnormal{d}\omega\ S(\mathbf{q},\omega)\ \omega^\alpha\,.\label{eq:define_moments}
\end{equation}
The zeroth frequency moment is simply the DSF normalization given by the static structure factor $M_S^{(0)}(\mathbf{q})=S(\mathbf{q})=F(\mathbf{q},0)$. Odd frequency moments of arbitrary order can, in principle, be computed from iterated commutators~\cite{quantum_theory,tolias_cpp_2026}, while a series expansion of even frequency moments in terms of odd frequency moments was recently derived by Tolias \textit{et al.}~\cite{Tolias_POP_2025}. However, closed form expressions are explicitly known only for the first and third frequency moments. The first frequency moment is given by the universal f-sum rule, $M^{(1)}_S(\mathbf{q})=-q^2/2$~\cite{pines_nozieres_bookI,quantum_theory}, while the third frequency moment or cubic sum rule can be straightforwardly rewritten as~\cite{PuffSumRule,quantum_theory} 
\begin{equation}\label{eq:K}
    K = \frac{1}{q^4}M^{(3)}_S(q)-\frac{1}{8}q^2-\frac{2\pi n}{q^2}+\frac{\omega_\textnormal{p}^2}{2q^2}I(q)\ ,
\end{equation}
with $\omega_\textnormal{p}=\sqrt{3/r_s^3}$ the plasma frequency, $n$ the number density, and $I(\mathbf{q})$ the Pathak-Vashishta functional of $S(\mathbf{q})$ whose integral form reads as\,\cite{PathakVashishtaScheme}
\begin{align}
    I(q) &= \frac{1}{8\pi^2 n}\int_0^\infty \textnormal{d}k\ k^2 \left[1-S(k)\right]\times\left[
\frac{5}{3} - \frac{k^2}{q^2} +\right.\\\nonumber &\quad \left. \frac{\left(k^2-q^2\right)^2}{2kq^3}\textnormal{ln}\left|\frac{k+q}{k-q}\right|
    \right]\,.\label{eq:PV}
\end{align}
We stress that the rhs of Eq.~(\ref{eq:K}) must not depend on $q$, even though each individual term clearly does.

Combining Eqs.(\ref{eq:Laplace},\ref{eq:define_moments}), it is straightforward to derive~\cite{Dornheim_moments_2023}
\begin{equation}
    M_S^{(\alpha)}(\mathbf{q})=(-1)^\alpha
    \frac{\partial^\alpha}{\partial\tau^\alpha}F(\mathbf{q},\tau)\Big|_{\tau=0}\,, \label{eq:define_derivative}
\end{equation}
that enables direct extraction of the frequency moments of the DSF from the ITCF, which is readily available from PIMC simulations. Despite the circumvention of the analytic continuation, the computation of the ITCF derivatives requires some additional considerations; PIMC yields $F(\mathbf{q},\tau)$ on a discrete equidistant $\tau$-grid of $P\sim10^2$ size, making the direct evaluation of numerical derivatives challenging. We follow the suggestion of Dornheim \textit{et al.}~\cite{Dornheim_moments_2023} and express the ITCF as a Taylor expansion around $\tau=0$,
\begin{eqnarray}\label{eq:I_like_toothpaste}
    F(\mathbf{q},\tau) = \sum_{\alpha=0}^\infty\underbrace{\frac{1}{\alpha !} \frac{\partial^\alpha}{\partial\tau^\alpha}F(\mathbf{q},\tau)\Big|_{\tau=0}}_{c_\alpha(\mathbf{q})}\   \tau^\alpha = \sum_{\alpha=0}^\infty c_\alpha(\mathbf{q})\tau^\alpha\, ;
\end{eqnarray}
a canonical polynomial fit to the PIMC results then gives the sought after frequency moments from the fitting coefficients $c_\alpha(\mathbf{q})$ via $M_S^{(\alpha)}(\mathbf{q}) = (-1)^\alpha c_\alpha(\mathbf{q}) \alpha!$.

\begin{figure}[t!]\centering
\includegraphics[width=0.485\textwidth]{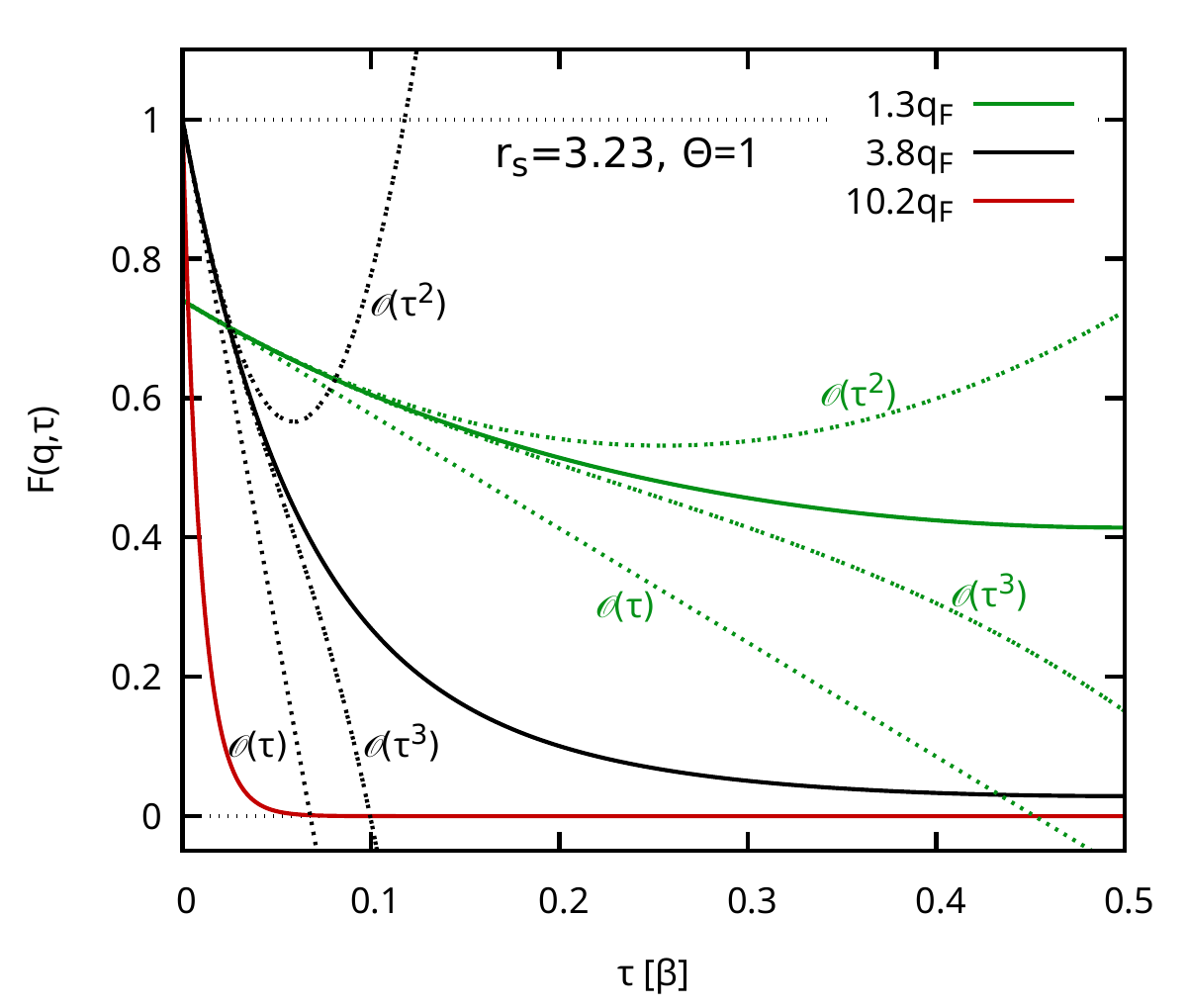}
\caption{\label{fig:ITCF} \emph{Ab initio} PIMC simulations of the warm dense paramagnetic UEG at $r_s=3.23$ and $\Theta=1.0$ for $N=4$ electrons with $P=10^3$ imaginary-time slices. Results for the ITCF dependence on the imaginary time at three wavenumbers. The dotted, double-dotted and triple-dotted lines correspond to the polynomial expansion of Eq.(\ref{eq:I_like_toothpaste}) evaluated up to first, second and third order, respectively.
}
\end{figure}

\section{PIMC results} Fig.\ref{fig:ITCF} features \emph{ab initio} PIMC results for the ITCF $F(\mathbf{q},\tau)$ of the UEG at the Wigner-Seitz radius $r_s=3.23$ and reduced temperature $\Theta=1/\beta E_\textnormal{F}=1$ (with $E_\textnormal{F}$ the Fermi energy) for $N=4$ unpolarized electrons. All the PIMC results have been obtained using the extended ensemble sampling scheme~\cite{Dornheim_PRB_nk_2021} as implemented in the \texttt{ISHTAR} code~\cite{ISHTAR}; they are freely available in an online repository~\cite{repo}. Such WDM conditions can be realized experimentally, e.g., with hydrogen jets, which in turn can be probed with XRTS~\cite{Zastrau,Fletcher_Frontiers_2022}. It is sufficient to restrict ourselves to the half range of $\tau\in[0,\beta/2]$ due to the symmetry $F(\mathbf{q},\tau)=F(\mathbf{q},\beta-\tau)$ that directly follows from the detailed balance of the DSF, $S(\mathbf{q},-\omega)=S(\mathbf{q},\omega) e^{-\beta\omega}$~\cite{Dornheim_T_2022}.
The colors distinguish three different values of the momentum transfer $q$, with the main trend being the increasing steepness of the $\tau$-decay; see Refs.~\cite{Dornheim_insight_2022,Dornheim_PTR_2022} for an extensive discussion.
The solid curves show the PIMC results and the dotted curves show the polynomial expansion Eq.(\ref{eq:I_like_toothpaste}) evaluated up to different $\tau$ order. Specifically, the difference between the double- and triple-dotted curves highlights the impact of the $\tau^3$ term, used to extract the cubic sum rule $M^{(3)}(\mathbf{q})$ and thus the kinetic energy $K$.

\begin{figure}[t!]\centering
\includegraphics[width=0.485\textwidth]{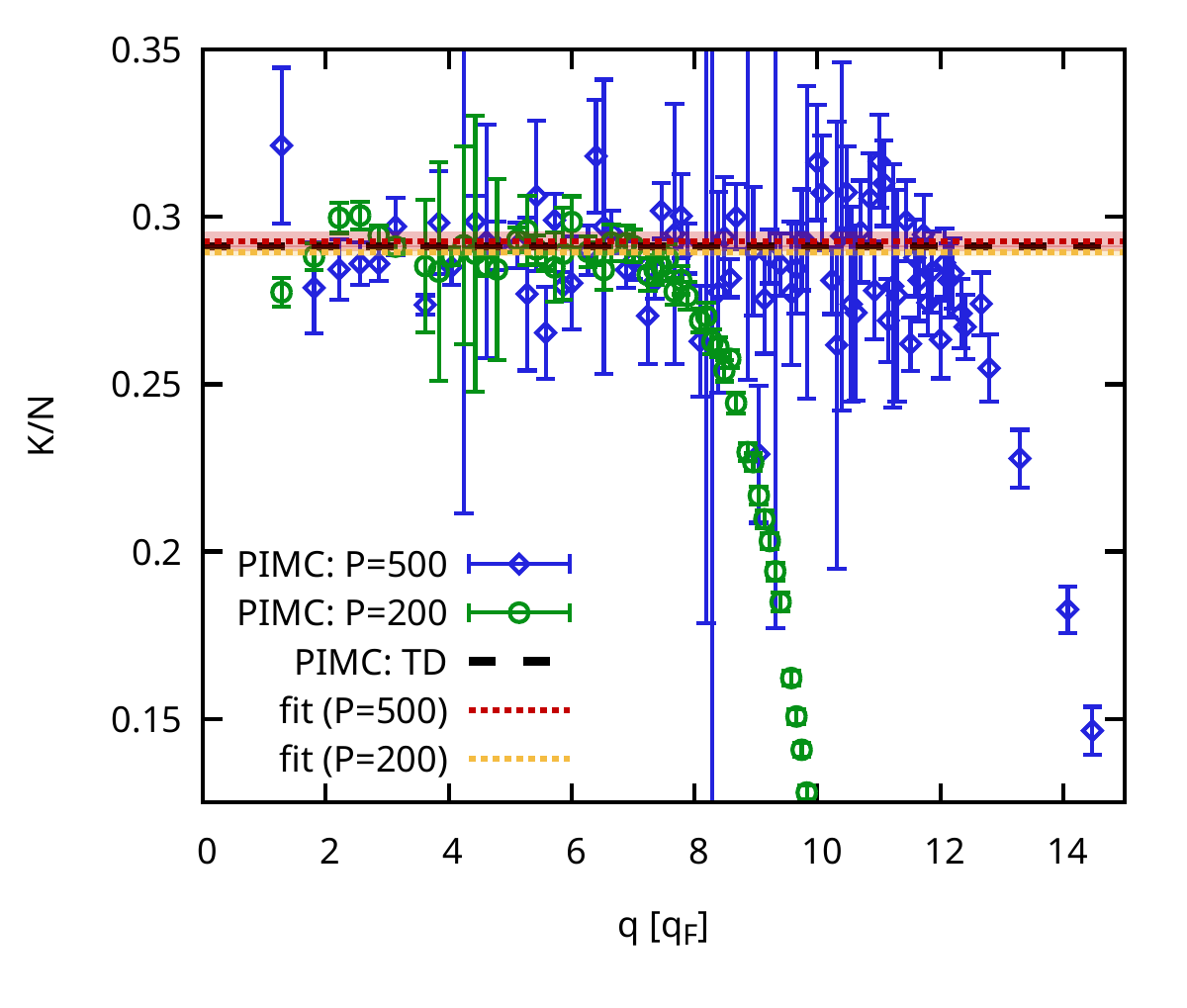}
\caption{\label{fig:compare} \emph{Ab initio} PIMC simulations of the warm dense UEG at $r_s=3.23$ and $\Theta=1.0$ for $N=4$ unpolarized electrons. The cubic sum rule route to the kinetic energy; computed from $M_S^{(3)}$ via Eq.~(\ref{eq:K}). Green circles and blue diamonds: PIMC results for $P=200$ and $P=500$ imaginary-time slices; dashed black line: thermodynamic estimator for $P=500$; dotted red line (shaded red area): fit to the $P=500$ PIMC data for $q\leq9q_\textnormal{F}$ (associated uncertainty); dotted yellow line (shaded yellow area): fit to the $P=200$ PIMC data for $q\leq6_\textnormal{F}$ (associated uncertainty).}
\end{figure} 

The latter is shown in Fig.\ref{fig:compare} for the same UEG parameters with the green circles and blue diamonds obtained for $P=200$ and $P=500$ imaginary-time slices, respectively. In addition, the result from the thermodynamic estimator (evaluated for $P=500$) has been included as the dashed horizontal line for comparison. First, it is straightforward that the cubic sum rule $K$ results are in very good agreement with the thermodynamic $K$ up to some maximum $q$ value that depends on $P$. This is easy to understand, as $F(\mathbf{q},\tau)$ becomes increasingly steep with increasing $q$, thus requiring a denser $\tau$-grid to obtain a meaningful polynomial fit. Therefore, results obtained with $P=500$ should remain stable up to larger $q$ values compared with results obtained with $P=200$, as confirmed. Second, fits to the $P=200$ and $P=500$ datasets for $q\leq6q_\textnormal{F}$ and $q\leq9q_\textnormal{F}$ respectively, see the dotted lines, led to $K_{200}=0.2895(13)$ and $K_{500}=0.292(3)$, to be compared with the thermodynamic estimator $K_{\mathrm{TD}}=0.2912(2)$ value. Results highlight the consistency of the PIMC method, but also indicate the limited utility of Eq.(\ref{eq:K}) as an alternative kinetic energy PIMC estimator~\cite{Janke_JCP_1997}. More PIMC results are discussed in the Supplemental Material~\cite{supplement}.

\section{Dielectric results} We proceed with approximate results for $S(\mathbf{q},\omega)$ from different dielectric schemes and we examine whether they allow for a consistent extraction of the kinetic energy from the cubic sum rule.  Recall that, at finite temperatures, it is more convenient to numerically solve dielectric schemes at the Matsubara frequency domain~\cite{stls,Tolias_PRB_2024} which leads to the Matsubara local field correction (LFC) $G(\mathbf{q},\ell)$ and then the Matsubara density response $\chi(\mathbf{q},\ell)$. This makes the ITCF immediately accessible through its Fourier-Matsubara expansion~\cite{tolias2024fouriermatsubara}. Unfortunately, a term-by-term differentiation of this Fourier series is not permissible at $\tau=0$, thus blocking the path to the frequency moments via Eq.(\ref{eq:define_derivative}) and necessitating their evaluation via the Eq.(\ref{eq:define_moments}) definition, which requires knowledge of $S(\mathbf{q},\omega)$. This is possible without analytic continuation by utilizing the set of equations of the dielectric formalism in the real frequency domain. In particular, the converged SSF $S(\mathbf{q})$ is used to calculate the 
LFC via the specific dielectric closure functional, which is used to evaluate the dynamic complex density response via the  polarization approach expression, whose imaginary part is utilized to compute $S(\mathbf{q},\omega)$ via the fluctuation-dissipation theorem~\cite{stls}. The interested reader is referred to the Supplemental Material for details\,\cite{supplement}. Focus lies on dielectric schemes that feature a static local field correction, this being a proof-of-principle study.

\begin{figure}[t!]\centering
\includegraphics[width=0.425\textwidth]{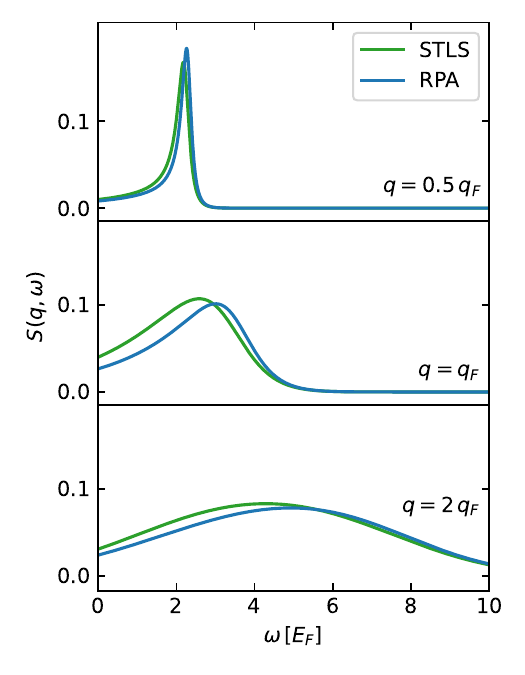}
\caption{\label{fig:fotis1} Dynamic structure factor $S(\mathbf{q},\omega)$ of the unpolarized UEG at $r_s=3.23$ and $\Theta=1$ obtained from dielectric theories. STLS: solid green line, RPA: solid blue line. Results for 3 wave-numbers: $q=0.5q_F$ (top), $q=q_F$ (mid), $q=2q_F$ (bottom).}
\end{figure} 

\begin{figure}[t!]\centering
\includegraphics[width=0.415\textwidth]{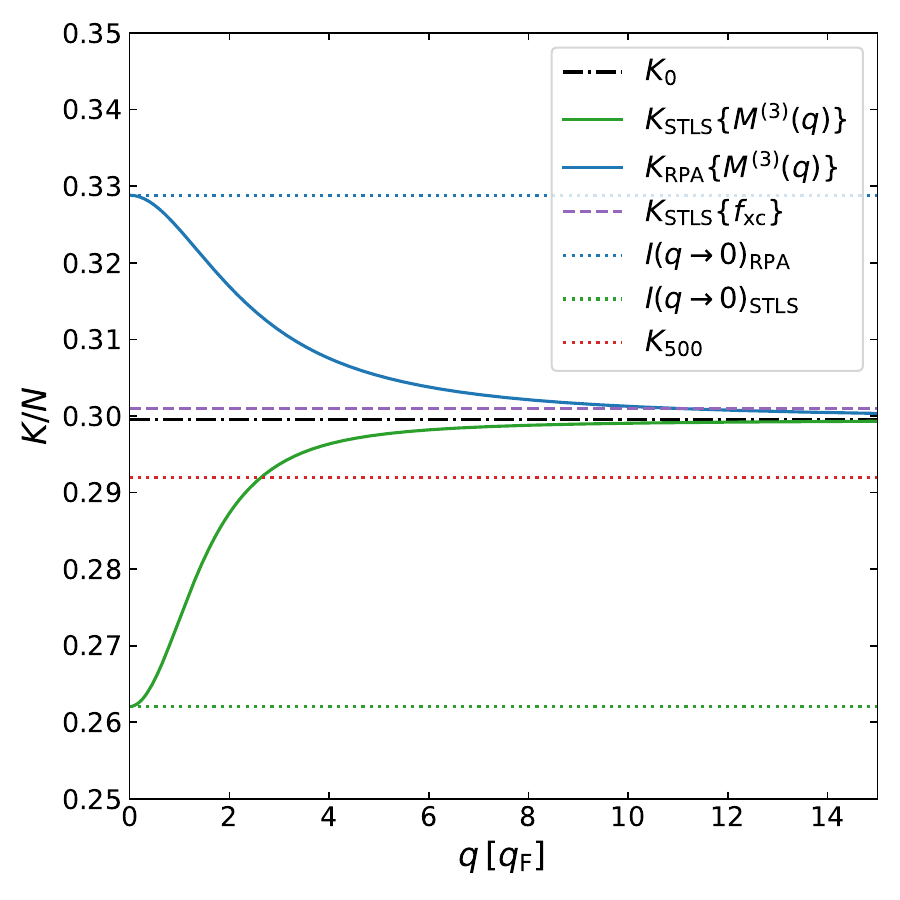}
\caption{\label{fig:fotis2} Total kinetic energy per electron of the unpolarized UEG at $r_s=3.23$ and $\Theta=1$ evaluated from dielectric schemes. $K_{\text{STLS}}\{M^{(3)}(q)\}$ (green solid line) and $K_{\text{RPA}}\{M^{(3)}(q)\}$ (blue solid line) denote the cubic sum rule result from Eq.(\ref{eq:K}); $K_0$ (black dot-dashed line) denotes the ideal kinetic energy. $I(q\to0)_{\text{RPA}}$ (blue dotted line) and $I(q\to0)_{\text{STLS}}$ (green dotted line) denote the exact long-wavelength limit; $K_{\text{STLS}}\{f_{xc}\}$ (purple dashed line) denotes the STLS kinetic energy from thermodynamic integration; $K_{500}$ (red dotted line) denotes the PIMC extracted kinetic energy from the cubic sum rule for $P=500$ imaginary time slices.}
\end{figure}

In Fig.~\ref{fig:fotis1}, we show $S(\mathbf{q},\omega)$ as obtained from the RPA and the STLS scheme for three representative wave numbers. Both approximations exhibit the expected dominant qualitative trends, with collective plasmon excitations at small wave numbers and an increasingly broadened response at higher wavenumbers $q$. Inclusion of a static LFC $G(\mathbf{q},0)$ in STLS leads to visible deviations from RPA, reflecting the impact of exchange-correlation effects on the density response. The limitations of the RPA and STLS become apparent when evaluating $K$ with the cubic sum rule. Fig.\ref{fig:fotis2} features the kinetic energy obtained by taking the third moment of the respective dielectric results for $S(\mathbf{q},\omega)$, by evaluating the Pathak-Vashishta functional with the respective dielectric results for $S(\mathbf{q})$ and by inserting them into Eq.(\ref{eq:K}). In contrast to the PIMC results discussed previously, the RPA (solid blue) and STLS (solid green) curves do not yield a $q$-independent estimate of the interacting kinetic energy. Instead, both schemes approach the ideal kinetic energy $K_0$ in the short-wavelength limit. This behavior is not accidental, but follows directly from the large-$q$ structure of the exact dynamic $G(\mathbf{q},\omega)$ entering the third-moment sum rule. 

To better illustrate this, we decompose the kinetic energy as $K=K_0+K_{xc}$, with $K_{xc}$ the XC contribution that can be isolated from the cubic sum rule as~\cite{dornheim2024MatsubaraResponse,ToliasqSTLS}
\begin{equation}
    \label{eq:K_XC}
    K_{xc}=\frac{\omega_p^2}{2q^2} \Big[ I(\mathbf{q}) -\lim_{\omega \to \infty} G(\mathbf{q},\omega)\Big]\,.  
\end{equation}
Static LFC approximations such as the RPA and STLS do not contain frequency constraints. In RPA, $G(\mathbf{q},\omega)=0$ by definition, while in STLS the static $G(\mathbf{q})$ approaches a finite constant in the short-wavelength limit and scales as $\propto{q}^2$ in the long-wavelength limit\,\cite{ToliasqSTLS}. Additionally, $I(\mathbf{q})$ is finite for $q\to\infty$ and scales as $\propto{q}^2$ for $q\to0$. As a result, the bracketed quantity in Eq.(\ref{eq:K_XC}) remains bounded in the large $q$ limit, and the explicit pre-factor $1/q^2$ forces $K_{xc}\to 0$, thus reducing the cubic sum rule extracted kinetic energy to the ideal contribution $K_0$. In the opposite long-wavelength limit, the scaling of $I(\mathbf{q})$ and $G(\mathbf{q})$ as $q^2$ cancels the prefactor so that Eq.~\ref{eq:K_XC} approaches a finite constant. The observed convergence of both RPA and STLS to the ideal kinetic energy at large $q$ should not be interpreted as a prediction for the interacting system. Rather, it reflects the strong violation of the third moment sum rule by these schemes. Finally, for comparison, Fig.~\ref{fig:fotis2} includes the kinetic energy obtained from STLS via the thermodynamic derivatives of the exchange-correlation free energy $f_{\mathrm{xc}}$. This further illustrates that the two kinetic energy routes are not consistent for these two static LFC approximations, as opposed to PIMC, which yields self-consistent results as expected. More dielectric results are discussed in the Supplemental Material~\cite{supplement}.

\section{Discussion} Herein, we explored the possibility of kinetic energy extraction from the third frequency moment of the dynamic structure factor. It is emphasized that the availability of an explicit expression which connects the expectation value of a single-particle operator ($\hat{K}$) to a two-body correlation function is rather remarkable and contrasts $\mathbf{q}$-integral expressions for the interaction energy in terms of $S(\mathbf{q},\omega)$ and for the kinetic energy in terms of the single-particle spectral function $A(\mathbf{q},\omega)$~\cite{hamann2026abinitiopathintegralmonte,Kraeft_PRE_2002}. Quasi-exact PIMC results for the ITCF $F(\mathbf{q},\tau)$ were utilized to extract $M^{(3)}(\mathbf{q})$ and $K$, which confirmed the expected self-consistency but also revealed limitations of the cubic sum rule PIMC estimator compared to the traditional thermodynamic PIMC estimator. Approximate dielectric results for the DSF $S(\mathbf{q},\omega)$ with the RPA and STLS schemes were utilized to extract $M^{(3)}(\mathbf{q})$ and $K$, which revealed that the third frequency moment constraint is violated with the extracted $K$ depending strongly on the wavenumber and tending to the ideal value $K_0$ in the short-wavelength limit due to the bound asymptote of the static $G(\mathbf{q})$. The above results demonstrate the practical viability of the approach and emphasize its potential value for future applications.

From a theoretical perspective, Eq.(\ref{eq:K}) can prove valuable as a check of the self-consistency of different methods, and as an additional constraint for the development of improved models and closure relations -- this concerns fully quantum dielectric schemes that feature a dynamic local field correction, and, in particular, possible extensions of the remarkably accurate qIET approach to include viscoelastic effects near Wigner crystallization\,\cite{Tolias_JCP_2021,Tolias_JCP_2023}. This also concerns linear-response time-dependent DFT calculations of $S(\mathbf{q},\omega)$~\cite{moldabekov2025applying,Moldabekov_MRE_2026}, where Eq.~(\ref{eq:K}) can be used to (1) benchmark XC-functionals and (both static and dynamic) XC-kernels, (2) estimate the full kinetic energy including the notoriously elusive XC-contribution separately from the full energy of the system. Future efforts might also include the generalization of the cubic sum rule to multi-component systems, which, in turn, will allow us to check the consistency of ubiquitous WDM models such as the aforementioned Chihara decomposition~\cite{Chihara_1987,bellenbaum2026xraydiagnosticsanalysisverification,boehme2023evidence}.

An even more ambitious project involves the utilization of the multi-component generalization of Eq.~(\ref{eq:K}) for the extraction of the kinetic energy from XRTS experiments. To this end, one might combine XRTS measurements at one or two scattering angles (and, thus, $\mathbf{q}$-values) with x-ray diffraction measurements to obtain the static structure factor $S(\mathbf{q})$, needed to evaluate the structural contribution $I(\mathbf{q})$, over a sufficient range of momentum transfers. Such an analysis will likely be carried out in the imaginary-time domain, where the two-sided Laplace transform facilitates a straightforward deconvolution of the relevant physics information~\cite{gawne2026modelfreeinterpretationxraythomson}. The key challenge will be to resolve the scattering signal over a sufficiently broad range of frequencies to achieve convergence that is already challenging for the first moment~\cite{Dornheim_SciRep_2024,Dornheim_NatComm_2025}.
In this regard, our idea may be fielded at modern x-ray free electron laser facilities, which provide high-brilliance, high-repetition rate x-ray pulses to probe WDM produced in self-scattering experiments~\cite{kraus_xrts, Sperling_PRL_2015}, short-pulse-laser-driven matter~\cite{Fletcher2015,Fletcher_Frontiers_2022}, and, recently, high-repetition rate long pulse lasers for shock compression experiments~\cite{DiPOLE100X,Bespalov2026}.


Finally, we note that the present analysis is rather general and extends to the full gamut of quantum many-body systems beyond warm dense matter; a case in point is given by ultracold atoms, (1) which are routinely simulated with PIMC methods~\cite{Filinov_PRA_2012,Ferre_PRB_2016,Dornheim_SciRep_2022,Boninsegni1996,cep,Saccani_Supersolid_PRL_2012} (and can be described with dielectric theories); (2) where the DSF can be measured experimentally with different techniques~\cite{Landig2015}.


\acknowledgments
\section{Acknowledgements} This work has received funding from the European Research Council (ERC) under the European Union’s Horizon 2022 research and innovation programme (Grant agreement No. 101076233, "PREXTREME"). 
Views and opinions expressed are however those of the authors only and do not necessarily reflect those of the European Union or European Research Council Executive Agency. Neither the European Union nor the granting authority can be held responsible for them. This work has received funding from the European Union's Just Transition Fund (JTF) within the project \emph{R\"ontgenlaser-Optimierung der Laserfusion} (ROLF), contract number 5086999001, co-financed by the Saxon state government out of the State budget approved by the Saxon State Parliament. This work has received funding from the German Federal
Ministry of Research, Technology and Space (BMFTR) via the
ErUM Data project “DEMOS” (Grant No. 05D25CR1). T.D.~gratefully acknowledges funding from the Deutsche Forschungsgemeinschaft (DFG) via project DO 2670/1-1. M.P.B.'s work was performed under the auspices of the U.S. Department of Energy by Lawrence Livermore National Laboratory under Contract No. DE-AC52-07NA27344. M.P.B. was supported by the Laboratory Directed Research and Development (LDRD) Grant No.~25-ERD-047.

The computations were performed on a Bull Cluster at the Center for Information Services and High-Performance Computing (ZIH) at Technische Universit\"at Dresden and at the Norddeutscher Verbund f\"ur Hoch- und H\"ochstleistungsrechnen (HLRN) under grant mvp00024. The authors gratefully acknowledge the computing time made available to them on the high-performance computer Otus at the NHR Center Paderborn Center for Parallel Computing (PC2) (ID 27589, HiFi-WDM). This center is jointly supported by the Federal Ministry of Research, Technology and Space and the state governments participating in the National High-Performance Computing (NHR) joint funding program.

\bibliographystyle{iopart-num}
\bibliography{bibliography2.bib}

\end{document}